\documentclass{article}
\usepackage{amsfonts}
\newcommand{\beq}{\begin{equation}}
\newcommand{\eeq}{\end{equation}}
\begin{document}

\title{\bf Gas  condensation within a bundle of carbon nanotubes -- effects of screening.}

\author{M.K.Kostov, J.C.Lewis and M.W.Cole\\
{\em Department of Physics,}\\
{\em The Pennsylvania State University, University Park, PA 16802, USA}}

\maketitle



One of the principal themes of condensed matter physics is the sensitivity of a system's behavior to its dimensionality \cite{lieb,tak,book}. Much interest in surface physics is attributable to the rich diversity of possible two dimensional ($2$D) phase diagrams. The discovery of the carbon nanotube (NT) has created the possibilty of interesting $1$D and quasi - $1$D phases involving adsorbed atoms and molecules[$4$-$9$]. In the case of small molecules/atoms, such as H$_2$, He and Ne, the most energetically favored site is believed to be the interstitial channel (IC) between NTs \cite{stan}, which is indeed a $1$D domain. In this paper we evaluate the phase behavior of a system of many H$_2$ molecules within the ICs. An important role is played by the carbon environment, which screens the attraction between hydrogens. As a result, we find that  the ground state cohesive energy of the system is reduced to a very small value ( less than $5\%$ of the well depth of the free space pair potential ). In this paper, we estimate the critical temperature for the condensation of hydrogen using alternatively   localized and delocalized models of the hydrogen. A comparison is presented with similar results for the adsorption of neon and helium. In all cases, the effects of screening in reducing the transition temperature are significant.

Our study is a successor to a recent investigation of the intrachannel screening \cite{kost}. That project evaluated the Axilrod-Teller-Muto triple dipole (DDD) interaction between two hydrogen molecules within the same IC and the surrounding carbon atoms. The DDD interaction has been shown empirically to account quantitatively for the three-body interaction involving simple gases in $3$D and semiquantitatively with similar effects in adsorbed films \cite{sayk,bark}. The resulting net interaction is:
\beq
V_{total}(\vec{r}) = V(\vec{r}) + V_{DDD}(\vec{r})  \, ,
\eeq
where $V(\vec{r})$ is the free space interaction. $V_{total}(\vec{r})$ 
has a well depth, $1.35$ meV, which is just $46\%$ of the free space well depth. As a consequence of the reduced attraction the H$_2$ dimer's binding energy falls from $4.5$ K to zero. This absence of the dimer bound state implies that the ground state of the strictly $1$D system is a gas \cite{krotch}. This result is manifested in the absence of a finite $\rho$ minimum in the $1$D equation of state $E_{1D}(\rho)$ at temperature $T=0$\, in Fig. $1$ (dashed curve); here $\rho$ is the $1$D density. This curve has been obtained by Boronat and Gordillo \cite{bor} using their diffusion Monte Carlo method, together with the screened interaction of Kostov {\em et al.} \cite{kost}. The curve contrasts strikingly with the result of a previous study which found a liquid with cohesive energy $4.83$ K for the $1$D system and the free space interaction between H$_2$ molecules \footnote{Because of some uncertainty about the adequacy of the DDD in describing the screening, we have explored the consequence of artificially reducing this interaction. Multiplying this interaction  by $0.845 \pm 0.01$ yields  the threshold for there to be a dimer bound state. Multiplication by one-half yields a dimer binding energy of $1.1$ K .}.

To study the collective properties of the H$_2$, we must take into account the interaction between molecules of neighboring channels. Fig. $2$ depicts both the unscreened and the screened intermolecular interactions, the latter including the DDD term which we evaluated using the same procedure as Kostov {\em et al.} \cite{kost}. Similarly, Fig. $3$ and Fig. $4$ illustrate the DDD screening effect in the interchannel interaction for the case of He and Ne atoms, adsorbed within the ICs. In all three cases, we observe quite a large DDD effect, which alters significantly the free space interchannel potential. Note that the distance scale in Fig. $2$, Fig. $3$ and Fig. $4$ is the longitudinal (along the axis of the IC) displacement $z$ between the two particles of neighboring channels. The spacing between neighboring ICs is a fixed parameter $d = 9.8$ \AA. 

One can easily obtain a rigorous upper bound for the interacting system's energy. To do so, we identify the system of uncoupled H$_2$ (i.e. independent ICs) as the unperturbed system. The ground state of this latter system is a set of uncondensed $1$D gases. The variational theorem asserts that the fully interacting system's energy satisfies

\beq 
E(\rho) \le E_{var}(\rho) \, ,
\eeq
\beq
E_{var}(\rho) = E_{1D}(\rho)+\frac{3\rho}{2}\int_{-\infty}^{\infty}\mathrm{d}z \,V_{total}\Big[(z^{2}+\rho^{2})^{1/2}] \,.
\eeq

The prefactor of $3/2$ is derived from a coordination number of $3$ neighboring ICs and the need to avoid double counting. More distant tubes (omitted here) would further lower this variational energy by less than $1\%$, which we neglect. Note that in Fig. $2$ the DDD term has regions which are both positive and negative. The integral is such that the negative contribution outweighs the positive contribution. This means that the environment {\em enhances} the interaction between molecules in adjacent channels. This might be called an ``antiscreening effect''.

Fig. $1$ presents the resulting variational energy. There is a minimum at density $\rho_{0}=0.091$ \AA$^{-1}$, at which the ground state energy per particle is $0.15 \pm 0.01$ K . This is a remarkably low density and weakly bound liquid. It is analogous to that found recently for $^4$He in the same environment \cite{col}. The latter calculation does {\em not} include environmental screening however.

What is the phase behavior of this unusual system?
As was found for $^4$He, there is a low temperature region of separation into liquid and vapor phases. A droplet of ``liquid'' will involve  condensation in contiguous ICs; this phase will be strongly anisotropic, and for that reason alone is worth further study. The critical region should be $3$D Ising-like, since the transition is inherently $3$D. The critical temperature may be estimated in at least two ways, based on alternative assumptions. The first of these ways is  simply to note the general similarity between cohesive energy and critical temperature. For $^4$He in $3$D for example, the cohesive energy is $E_{c} = 7.17$ K and the critical temperature is $5.2$ K. In $2$D, these values are also similar ($E_{c}= 0.85$ K, $T_{c}=0.87$ K)\cite{gor}. On this crude basis, we estimate $T_{c} \sim 0.1$ K for H$_2$ in NTs.

However, a quite different estimate is obtained if the H$_2$ molecules become localized at sites created by the periodic potential provided by the NTs. In this case, a lattice gas model is appropriate. To describe the phase behavior of H$_2$ in this model we follow the procedure of Cole  {\em et al.} \cite{col}. In that approach we obtain $J_t \approx 14$ mK and $c  \approx 0.007$, where an anistropic simple cubic Ising model was assumed with an interaction strength $J_z$ between neighboring spins along the $z$ axis (within the same channel) and a transverse interaction $J_t = c J_z$. The critical temperature of the H$_2$ condensation transition in this lattice-gas model is then $T_c \sim 1.19$ K \cite{fish}.

Analogous assessments of the condensation properties of He and Ne in that model  yield  $J_t \approx 2.6$ mK, $ c  \approx 0.009$ and $T_c \sim 0.18$ K for He , and  $J_t \approx 15$ mK, $ c  \approx 0.009$ and $T_c \sim 1.05$ K for Ne. The previous study of He condensation found $T_c \sim 0.36$ K in the nonscreening case.

In summary, we have derived estimates of the transition temperatures for quasi-$1$D condensations of gases in nanotube bundles. The effect of the carbon environment is significant. It provides strong screening of the {\em intra}channel attraction but overall enhances the {\em inter}channel attraction \footnote{ If the screening of both intra- and interchannel interactions were omitted, we would find the following values of $T_c$ for H$_2$: $\sim 4.8$ K in the quasifree model and $\sim 3.01$ K in the localized model.}. These phase transitions would be dramatically observable in specific heat measurements at low temperature because the tubes are thermodynamically inert in that temperature regime.

We are grateful to M. Chan, P. Eklund, P. Sokol, V. Crespi and J.K. Johnson for helpful discussions. We express our thanks to J. Boronat and M.C. Gordillo for carrying out the diffusion Monte Carlo calculations described above. This research has been supported by the Army Research Office and the Petroleum Research Fund of the American Chemical Society.

\newpage

\begin{center}{\bf Figure Captions}
\end{center}

Fig. $1$. Energy per molecule of $1$D H$_2$ within a single channel (dashed curve from Ref.\cite{bor}) compared with the energy per molecule in the case when interchannel interaction is included (full curve).

Fig. $2$. Potential energy of interaction between two H$_2$ molecules in free space (dashed curve) and in adjacent channels (full curve) are shown. The difference arises from the DDD interaction of the two H$_2$ molecules and the surrounding carbon atoms (dotted curve). The abscissa is the difference in $z$ coordinates of the two molecules.

Fig. $3$. Same as Fig. $2$ but for He.

Fig. $4$. Same as Fig. $2$ but for Ne.

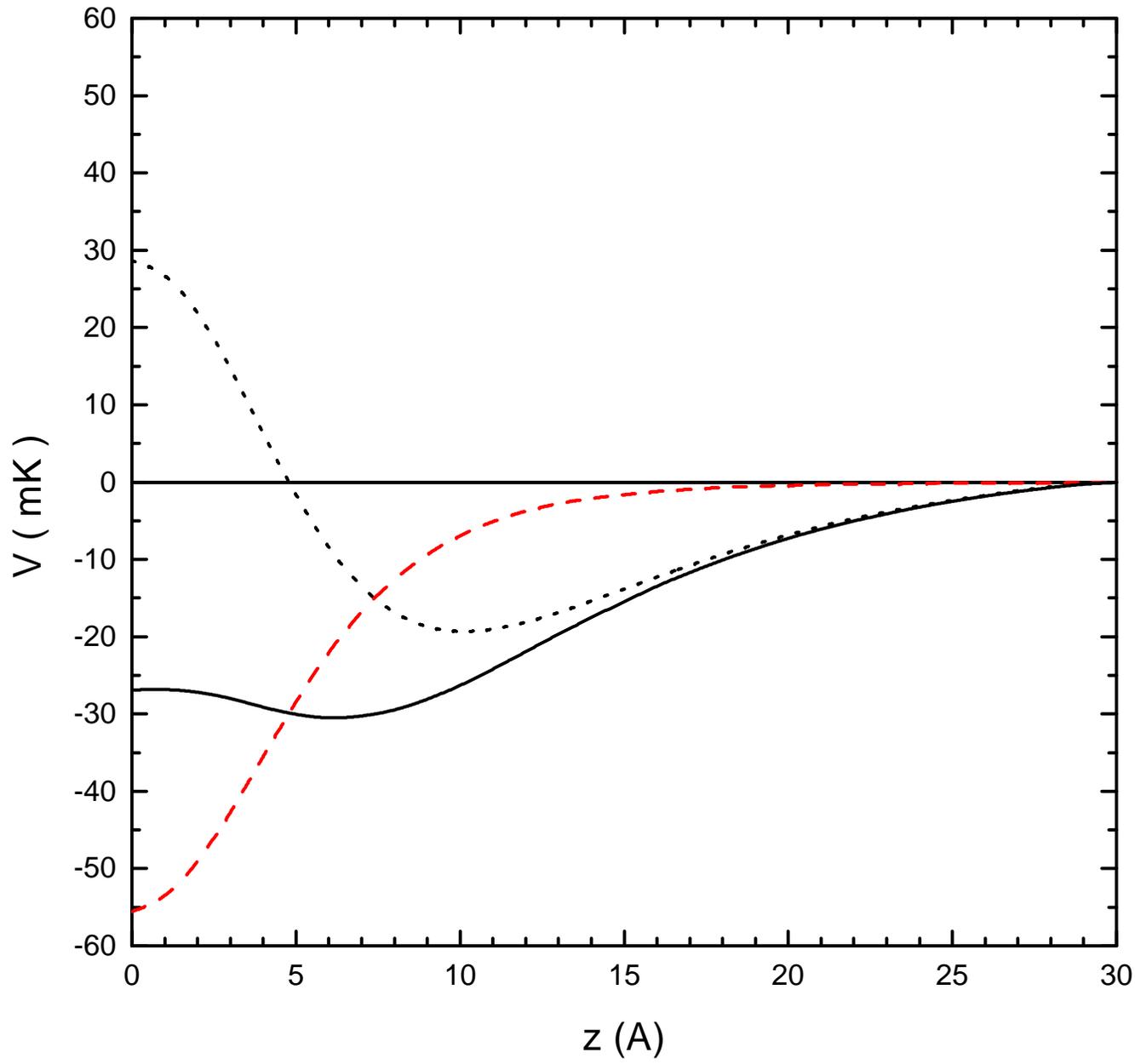
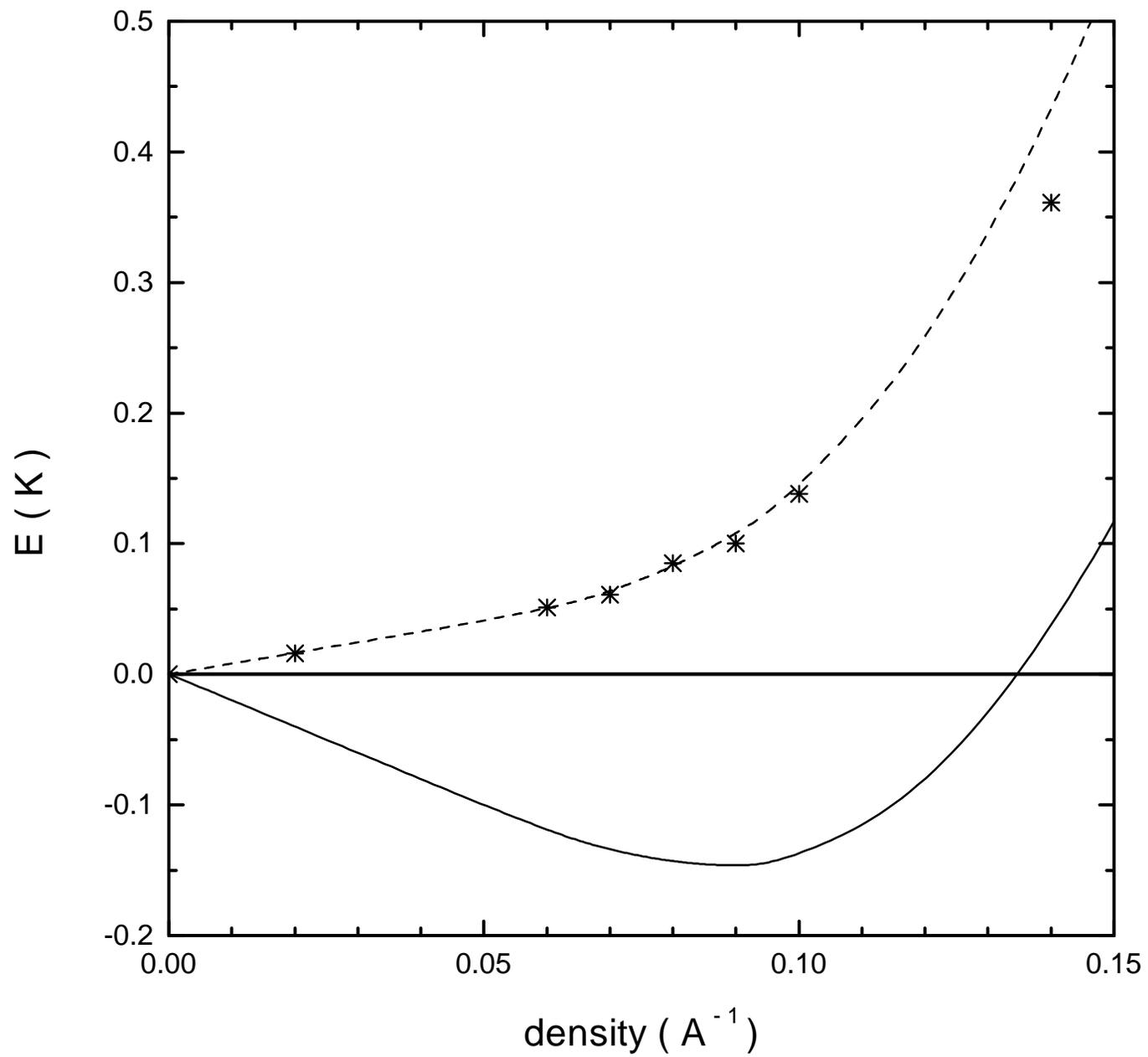
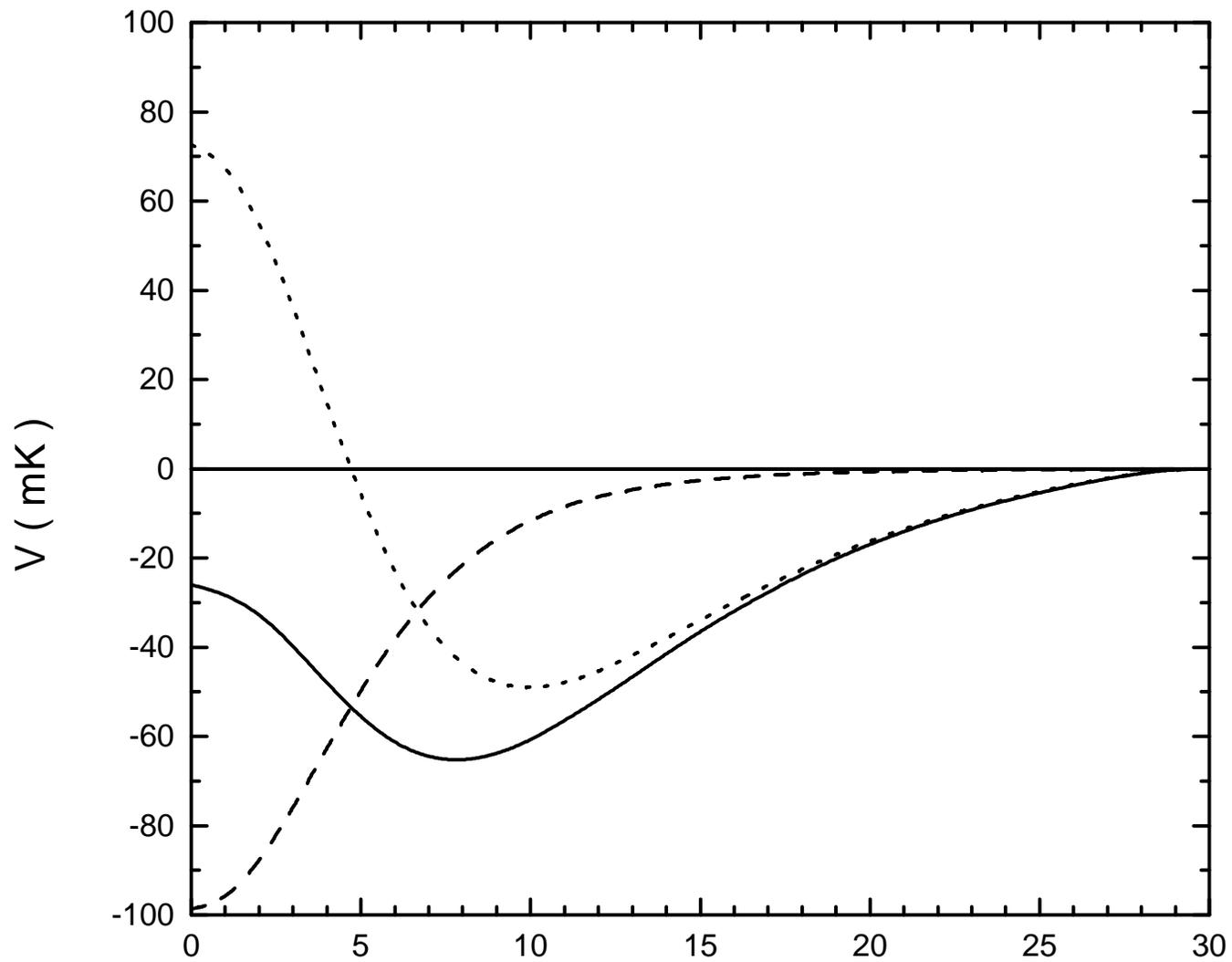
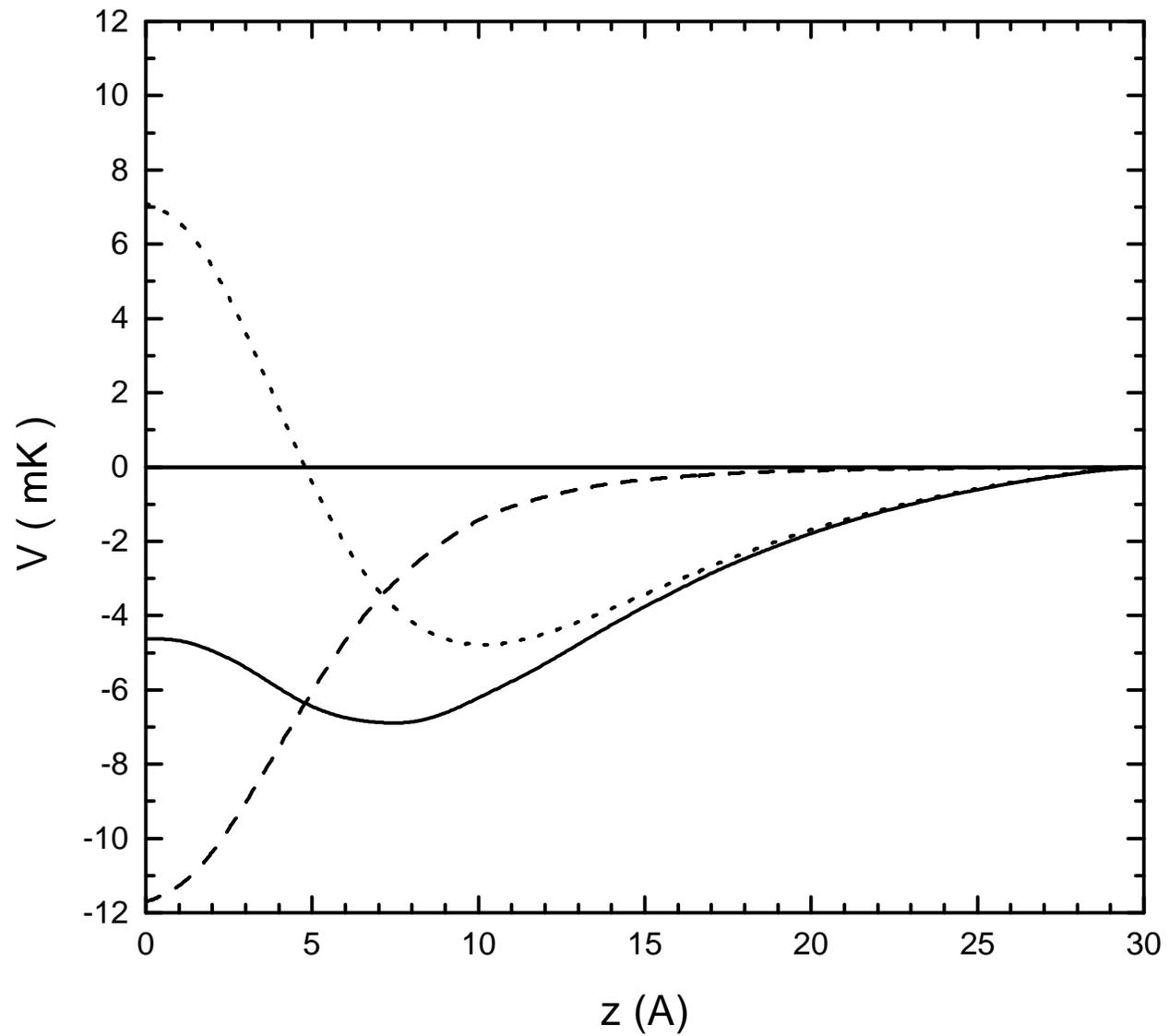
\end{document}